# The Added Mass of a Falling Coffee Filter

C. Boyle[1] and J. Pantaleone[1]

[1]Department of Physics, University of Alaska Anchorage, Anchorage, AK 99508

**Abstract**

Falling coffee filters are a common object of study in introductory physics laboratory classes. They are usually used to demonstrate the terminal velocity, and determine how the drag force depends on this velocity. However they can also be used to demonstrate other ways that the air affects the motion of the falling coffee filter. In particular, a qualitatively different fluid force acts during the initial acceleration from rest, and causes this acceleration to be much less than the acceleration of an object in vacuum. This fluid force, commonly known as the "added mass", is easily observable with coffee filters using either a sonic motion sensor or video analysis. Here the added mass is experimentally measured for a falling coffee filter. In addition, the theoretical value for the added mass of a coffee filter moving through an ideal fluid is calculated. The two values agree relatively well. Finally, it is shown how the added mass is relevant to experiments that only measure the terminal velocity.

## 1. Introduction

Falling coffee filters are a popular object of experimental study in the introductory science lab. There are many reasons for this popularity. Their motion is such that they can be studied with a variety of experimental means: sonic range finders, video analysis, timers, or even with only a meterstick. Also, they are inexpensive, readily available, and they can be stacked without changing their aerodynamic profile. In addition, they provide students with the opportunity to make log-log plots, in order to determine the power-law dependence of the steady-state drag force on the velocity [1,2]. Given this popularity, it is surprising that there is a basic aspect of their motion that has heretofore been ignored---the added mass.

For the motion of an object like a falling coffee filter (axisymmetric, with no change in orientation as it falls) the fluid dynamic forces acting on it are usually parameterized as [3,4]

$$\vec{F}_{drag} = -m_A \frac{d\vec{V}}{dt} - \frac{1}{2} C \rho \pi R^2 V \vec{V} \tag{1}$$

The second term in Eq. (1) is discussed in most introductory physics textbooks---it is the steady-state drag force. Here $\rho$ is the fluid density, $R$ is the cross-sectional radius of the object, $\vec{V}$ is the velocity, $V$ is the magnitude of the velocity, and $C$ is the dimensionless drag coefficient. The first term in Eq. (1) is not found in most introductory physics textbooks. Here $m_A$ is commonly known as the "added mass", or sometimes the "virtual mass". The added mass force is usually taken into account by rearranging Newton's second law so that the total inertia of the object moving through the fluid is the sum of the mass of the object plus the added mass.

The existence of the added mass force is easy to understand. When an object moves through a fluid at rest, the fluid must move around the object with speeds proportional to that of the object. Thus when the object accelerates, the fluid moving around it must also be accelerated. Hence the effective inertia of an object is larger when it moves through a fluid. The added mass is present even for an ideal fluid (inviscid and incompressible), while the steady-state drag force vanishes then (this is known as D'Alembert's paradox). The added mass is a large effect when low density objects change their velocity, such as for the motion of a ship in the water or a blimp in the air. The earliest experimental investigations of the added mass were by Chevalier Du Buat [5,6], but the study of fluid forces on accelerating objects is still an active area of research (see e.g. [7] and references therein). The added mass can easily be measured in the physics classroom by observing the motion of a beach ball [8,9], or a pendulum [10], or a falling coffee filter.

Newton's second law for a falling coffee filter is

$$m \frac{dV}{dt} = mg - F_{\text{drag}} \tag{2}$$

Here downward is taken to be the positive direction, $m$ is the mass of the coffee filter, $g$ is the acceleration of gravity in vacuum, and $F_{\text{drag}}$ is the magnitude of the drag force. This equation can be rewritten as

$$\frac{dV}{dt} = A\left[1 - \left(\frac{V}{V_T}\right)^2\right] \tag{3}$$

where Eq. (1) was used. The parameters in Eq. (3) are $V_T$, the terminal velocity, which is the velocity when the acceleration is zero

$$V_T = \sqrt{\frac{2mg}{C\rho\pi R^2}} \tag{4}$$

and $A$, the acceleration when the velocity is zero

$$A = \frac{m}{m + m_A} g \; . \tag{5}$$

Eqs. (4) and (5) assume that the volume of the coffee filter is zero, so a buoyant force has not been included in them. From Eq. (4) we see that the terminal velocity, $V_T$, is independent of the added mass, $m_A$. From Eq. (5) we see that $A$ is sensitive to the added mass, especially when the mass $m$ is small, but is independent of $C$. From measurements of $A$, $m$, and $g$, the added mass, $m_A$, can be determined.

Assuming that $m_A$ and $C$ in Eq. (1) are constant, then $A$ and $V_T$ are constants, and then Eq. (3) can be integrated twice to yield the position as a function of time for an object released from rest at $t=0$ with $y=0$,

$$y(t) = \frac{V_T^2}{A} \text{Ln}\left[\text{Cosh}\left[\frac{A}{V_T}t\right]\right]. \tag{6}$$

Equation (6) is valid even when the buoyant force is important, however then the expressions for $V_T$ and $A$ are slightly different from those given in Eqs. (4) and (5).

## 2. Measuring the added mass.

The coffee filters used in this experiment were the "8-12 cup, basket coffee filters" made of white paper. These are inexpensive and readily available in large quantities. They had a bottom diameter of 8.7±0.1 cm, a height of 4.7±0.3 cm, and a top diameter of 15.1±0.3cm =2R. The latter uncertainties are larger, reflecting the fact that the edges were fluted and also that these dimensions grew somewhat with the number of filters that were stacked. Each coffee filter had a mass of approximately $m = 0.86$ grams.

Sonic range finders and video analysis both provide measurements of position versus time, and either can be used to extract $A$ and $V_T$ from a single observation of a falling coffee filter. For the experimental results presented here, a PASCO Motion Sensor II sonic range finder was used. The coffee filters were released by hand about 0.2 m under the range finder, about 2 meters above the ground, and with the closed, smaller side down. The parameters $A$ and $V_T$ were determined from the position versus time data using PASCO's Capstone data collection and analysis software. The dependence of these parameters on the mass of the coffee filters was analyzed using the spreadsheet software Excel.

Fig. 1 shows data collected for a single coffee filter at a sampling rate of 100 Hz. This data can be analyzed in different ways to find the parameters $A$ and $V_T$. One method is to fit quadratic and linear curves to the initial and final parts of the data, respectively, and from this determine $A$ and $V_T$ separately. The light data points in Fig. 1 show the parts of the data that were fit in this way, and the dotted curves are the resulting fits, extended past the fitted points. This method gives the parameter values $A = 2.96$ m/s² and $V_T = 1.12$ m/s, respectively. Another analysis method applied to this data was to use the analytical solution given in Eq. (6), and to fit it to all of the data points between the left edge of the left light region and the right edge of the right light region. The solid curve in this plot shows this fit, however it is difficult to see because it is hidden behind the data points, except at the right edge of the plot. This analysis method gives $A = 1.6$ m/s² and $V_T = 1.06$ m/s. The number of significant figures in these values indicate their certainty from small variations in the fit procedure. Both methods find that the initial acceleration is far smaller than $g$. Thus the added mass of the coffee filter is considerably larger than the mass of a single coffee filter. This demonstrates that the added mass is important for understanding the motion of a falling coffee filter. However the two data analysis methods give quite different values for the initial acceleration, and hence the added mass.

The difference in the fit parameters from the two different methods means that, while Eq. (1) provides a much better description of the drag force than when the added mass is neglected, there are some additional, small, fluid dynamic effects beyond those contained in Eq. (1) with constant $C$ and $m_A$. This should not be surprising. Steady state measurements of $C$ over a wide range of velocities find that it is generally a function of the speed, $V$ (see e.g. [3]). Similarly, measurements of the added mass, $m_A$, find that it also can vary (see e.g. [7]). A quantitative description of these small fluid effects will not be pursued here. Instead the focus here will be on how to best present the concept of added mass in an introductory lab experiment.

The consistency of the added mass in a coffee filter experiment is easy to demonstrate. In introductory science labs, the usual procedure in coffee filter experiments is to measure the fall parameters as a function of the mass of the falling coffee filters. With that in mind, Eq. (5) can be written as

$$\frac{1}{A} = \frac{m_A}{g}\frac{1}{m} + \frac{1}{g} \tag{7}$$

Thus a plot of $1/A$ versus $1/m$ should yield a straight line if the added mass, $m_A$, is a constant. This was tested by doing experiments where from $N = 1$ to 5 nested coffee filters were dropped from rest, with a minimum of 10 measurements for each $N$. The average value of $A$ was found for each case, and the standard deviation of these measurements was used as the statistical uncertainty in the average value. These statistical variations probably come from irregularities in the release of the filters, from small motions in the surrounding air, and possibly from turbulence in the wake. This data is plotted in Fig. 2.

The data in Fig. 2 were obtained using the maximum possible sampling rate available on the sonic range finder, 250 Hz. At this high sampling rate it is difficult for the motion sensor to track the filters over a long distance, so no attempt was made to measure $V_T$ in this set of experiments. Only data points with speeds in the approximate range $0 < V < V_T/4$ were fit to a quadratic equation to determine $A$ (using $V_T$'s from other measurements). This corresponds to fall distances ranging from 1.6 cm to 3.2 cm for $N = 1$ and 5, respectively. With a cut-off of $V_T/4$, the steady-state drag force at the end of the fitted region is expected to be only $(1/4)^2$ its max value (see Eq. (3)), and thus this force should contribute an error to the measured value of $A$ of order $(1/4)^3$, which is less than 2%. This is below the statistical variations in the measured values of $A$. The measurements of $A$ are well fit by a line, suggesting that it is reasonable to take the added mass as a constant with this analysis method. The value of this constant was determined from the average value of $A$ for each $N$, and these values were combined using a weighted average to give an overall value of

$$m_A = 1.9 \pm 0.1 \, \text{grams} \tag{8}$$

where the uncertainty is only statistical. For this calculation the measured mass of the filters was used, along with the local value of $g = 9.82$ m/s$^2$. The theoretical prediction using Eqs. (7) and (8) is plotted as the dotted line in Fig. 2.

A second set of experiments were performed, with the goal of extracting both of the parameters $A$ and $V_T$ from the observation of a single coffee filter fall. Here a sampling rate of 40 Hz was chosen, which allowed the filters to be tracked easily over a distance of 2.2 meters. Positions near the end of the run, 1.5 m $< y <$ 1.8 m, were fit to a line in order to determine the terminal velocity, $V_T$. Data in the range 1.8 m $< y <$ 2.2 m was not used because in this range the filters were observed to slightly slow down as they approached the floor, for the larger $N$ cases. The measured values of $V_T$ were analyzed as usually done for falling coffee filter labs [1,2]. The steady state drag force was assumed to have the form $kV^n$, where $k$ is a constant and $n$ is the scaling exponent. Then at steady state motion $mg = kV_T^n$, and a plot of log(m) versus log($V_T$) gives $n$ as the slope [see Supplementary Material for plot]. The measured data gave $n = 1.93 \pm 0.04$. If instead the value $n = 2$ was assumed, in agreement with Eq. (1), then the data yield a drag coefficient of $C = 0.71 \pm 0.04$. To find the initial acceleration, $A$, the data was too sparse at this sampling rate to choose the upper cut-off of the data range using the velocity, as done previously. Instead, data in the range 0 $< y <$ 3 or 4 cm was fit to a quadratic to find the initial acceleration (with the larger fall distance used at larger $N$). These measurements of $A$ were also reasonably well fit by a line on a plot of $1/A$ versus $1/m$, and yielded a value of $m_A = 2.4 \pm 0.4$ grams. The larger statistical uncertainty here is probably due to the far fewer data points being fit at this lower sampling rate.

The second data set was also analyzed by fitting positions over the range 0.2 m $< y <$ 1.8 m with Eq. (6). This analysis method gives the values $m_A = 3.0 \pm 0.2$ grams and $n = 1.88 \pm 0.03$. This analysis method is probably not the best for determining a value for $m_A$ or $n$ since Eq. (6) assumes the parameters $m_A$ and $C$ are constants, while the data suggests something slightly different (as illustrated in Fig. 1). However the results of this fit, with Eq. (6), are very useful as a model of the average motion of the filter. This value for $m_A$ will be used in section 4 to make predictions about the motion of the filters under different circumstances. First, a physical explanation for why this value of $m_A$ is slightly higher than that in Eq. (8) is given in the following section.

### 3. Understanding the measurements.

The added mass of an object depends on its size and on the density of the air. Consequently it is common practice to define a dimensionless added mass parameter, $C_A$, that only depends on the geometry of the flow around the object. For comparison purposes we shall here use the largest cross-sectional radius of the coffee filter, $R$, and define the dimensionless added mass coefficient as

$$C_A = \frac{m_A}{\rho R^3} \tag{9}$$

For the experiments conducted here, the air pressure, temperature and humidity were measured and used to calculate an air density of $\rho = 1.157$ kg/m$^3$.

The measured value of $C_A$ calculated from the value in Eq. (8) is given in Table 1. For comparison, Table 1 also includes values of $C_A$ calculated for the flow of an ideal fluid. Values of $C_A$ for simple geometries (like a disk or a sphere) are in the published literature, but not for coffee filters, so a calculation was performed to find $C_A$ for the shape of the coffee filters used in this experiment.

For an object that starts from rest and accelerates through an ideal fluid, the flow is everywhere vorticity free. Then the flow can be described by a velocity potential that satisfies Laplace's equation. To calculate $C_A$ for a coffee filter, Laplace's equation was solved numerically using the software Mathematica. The coffee filter was assumed to have axial symmetry, to turn the calculation from 3D to 2D. The coffee filter was taken to be impermeable to fluid flow, so the boundary conditions for the velocity potential were that the gradient of the potential perpendicular to the surface equaled the component of the object's velocity in that direction, and that the potential vanished at infinity. From the velocity potential, the kinetic energy per unit volume in the fluid was found, and then this was integrated over the fluid volume around the filter to find the total kinetic energy of the fluid. This was taken to be $(1/2)m_A V^2$, to determine the added mass. The result of this calculation is given in Table 1. This calculated value is larger than that of a thin, flat disk, as is to be expected since the coffee filter encloses a larger volume of air behind it than a disk. The calculated size is about half that of a hollow sphere, which is reasonable since the length of the coffee filter along the flow is much less than that of a sphere.

The value of $C_A$ from the measurement in Eq. (8) is a little larger than the ideal fluid value, but in approximate agreement. This agreement is not too surprising since this measured value is found from when the falling coffee filter is moving its slowest. That is when the flow is most likely to be laminar everywhere, and so most likely to resemble that of an ideal fluid [11]. As the coffee filter falls, its speed increases, and so does the dimensionless Reynolds number $Re = \rho(2R)V/\mu$, where $\mu$ is the viscosity. At terminal velocity

$$Re_{\text{Terminal}} \approx 1.0 \times 10^4 \sqrt{N}, \qquad (10)$$

where $N$ is the number of coffee filters. At the Reynolds numbers of Eq. (10), observations of the flow around disks and spheres suggests that the laminar flow around the coffee filter will be separated from the rear surface of the filter and there will be a turbulent, low pressure wake immediately behind the falling filter [3,11]. This wake is important for understanding both the steady state drag force and the changes in the added mass. It is the pressure difference between the front and rear of the filter that is primarily responsible for the drag force at the Reynolds numbers in Eq. (10). Also, this wake increases the length of fluid moving behind the coffee filter, and so increases the volume of fluid in motion. This suggests a larger added mass value then, and so qualitatively explains why larger values of the added mass are measured when a wider range of the position versus time data are included in the parameter fits, such as the fit to Eq. (6) in Fig. 1.

### 4. Implications for classroom measurements of the scaling exponent, $n$.

The added mass causes the initial acceleration to be quite small, hence it takes much longer than naively expected for the falling coffee filters to reach terminal velocity. The distance the coffee filters need to fall from rest to 95% of terminal velocity is given by

$$y_{95\%} \approx \frac{V_T^2}{A} 1.16 \tag{11}$$
$$\approx 0.64\text{m} + (N-1)0.14\text{m}$$

Here Eqs. (4) through (6) were used, along with $m_A = 3.0$ grams, $n = 2$, $m \approx N*0.86$ grams and $V_T(N=1) = 1.1$ m/s [see Supplementary Material for details]. In a typical classroom, the maximum fall distance is usually limited by the height of the student to about 2 m. The amount of usable distance is less than this since when the coffee filters are near the floor they slow down slightly, with more slowing at the larger $N$'s. Thus in experiments with motion sensors or video analysis, it is not possible to use more than about $N \approx 5$ or 6 coffee filters in a classroom and still have a sizeable distance range over which the coffee filters move at near terminal velocity.

There are a couple of different ways to measure $n$ without motion sensors or video analysis. These methods typically neglect the initial period of accelerated motion, and take the average velocity to be the terminal velocity. Thus these methods underestimate the terminal velocity, and generally this underestimation is worse for large $N$, since then the coffee filters fall farther before reaching terminal velocity, as indicated by Eq. (11). Hence these method typically give a value of the scaling exponent, $n$, that is larger than the physical value.

One "low tech" method uses only a stopwatch, and measures the time, $t$, for 1 through $N$ coffee filters to fall a fixed distance. Then $n$ is found from the slope of a plot of log($N$) vs log($t$). The error in this method can be calculated at leading order by expanding Eq. (6) for large times, and keeping the first correction term. After some manipulation [see Supplementary Material for details], one gets

$$n \approx 2 + 4\ln(2)\langle N \rangle \frac{V_{T1}^2}{g\,y}$$
$$\approx 2 + \frac{1.02\text{m}}{y} \tag{12}$$

Here $y$ is the fixed distance the coffee filters fall, $\langle N \rangle$ is the average number of filters dropped (taken here to be 3), and $V_{T1}^2$ is the terminal velocity of 1 coffee filter (see below Eq. (11)). The error, ($n$-2), is huge for smaller fall distances, which is understandable since then the filters spend more of the fall distance approaching the steady-state motion. Surprisingly, the added mass cancels out at leading order, but it does enter in the higher order terms. Eq. (12) shows that, to achieve accurate results with this method, one needs very large fall distances. Thus these experiment must be done with filters dropped from a balcony, or in a stairwell. In such an experiment, students can also observe the small horizontal motions common with the fall of light objects moving at the Reynolds numbers in Eq. (10). Such horizontal drifting is associated with the shedding of vortices from the wake of the falling object, and can produce zig-zag or spiral motion in falling spheres [3].

One "no tech" method uses only a device for measuring lengths. One filter and $N$ stacked filters are dropped simultaneously, from different heights, and the height ratio is found for when they hit the ground simultaneously. The $N = 1$ case is kept at a fixed height, $y_1$, and the distance $y_N$ of the stacked case is measured for different $N$, then a plot of $\log(N)$ versus $\log(y_N)$ will yield $n$ from the slope. The error is this method can be calculated at leading order, as was done in Eq. (12) [see Supplementary Material for details].

$$n \approx 2 + 2\ln(2)\left[\left\langle N^{1/2}\right\rangle - \frac{m_A}{m_1}\left\langle N^{-1/2}\right\rangle\right]\frac{V_{T1}^2}{g\,y_1} \tag{13}$$

Here $m_1$ is the mass of 1 coffee filter, and the other parameters are as given below Eq. (12). Eq. (13) shows that there is a cancellation in the correction term, so that its size will be relatively small. Thus this method gives good results when used for relatively small fall distances. The numerical value of the error is shown in Fig. 3, where the exact form of Eq. (6) was used since the heights used in this method are generally smaller. For no added mass, the value of $n$ is larger than the assumed value, $n=2$, as expected from Eq. (13) and as found for the previous method. However the presence of the added mass greatly decreases the size of the error, as predicted by Eq. (13). The added mass slows down the motion of the $N = 1$ case, and since the $N = 1$ case is used as a reference, this has the effect of making the measured $n$ value smaller than expected. The accuracy of this method is surprising since the coffee filters are typically not at terminal velocity for most of their falling distance.

## 5. Discussion.

The added mass of a coffee filter, $m_A$, is more than twice the mass, $m$, of the filter. Thus this is a large fluid dynamic effect that is easy to measure in a classroom experiment using sonic range finders or video analysis. When measuring the initial acceleration, $A$, one should choose as small a range as reasonable near the start of the motion to fit with a parabola, in order to minimize velocity dependent fluid effects. Plotting the measured values as $1/A$ vs $1/m$ yields a line that demonstrates the consistency of the added mass effect, see Fig. 2. The value of the added mass measured in this experiment will be close to the value calculated for an ideal fluid, see Table 1.

Experiments to measure the scaling exponent, $n$, can also be done with a "low tech" method that uses only a stopwatch, and a "no tech" method that uses only a meter stick or tape measure. The "no tech" method is more accurate for short fall distances because of a partial cancellation between the effect of the added mass and the effect of neglecting the initial acceleration. However the error in the "low tech" method is easier to understand, and also the careful observer can see the effects of vortex shedding during the large fall distances necessary for this method.

The added mass can also be measured in experiments using a beach ball [8,9] or a pendulum [10]. These experiments have the advantage that the measured values are much closer to the ideal fluid value. However the coffee filter experiments have the advantages that the added mass and the terminal velocity can both be measured in the same position versus time observation, and also that the mass of the coffee filters can be easily varied.

Photographs of fluid flow around moving objects are a valuable addition to these introductory fluid flow experiments. They help to develop intuition about the nature of fluid flow. Reference 11 is an especially good source of such images, and pdf's of this book are freely available from the publisher's website. It contains images of steady state flow around a variety of objects, and of the flow patterns around accelerated objects.

A measurement of the added mass provides students with a much deeper understanding of fluid forces. Also, it introduces students to the concept of an effective mass for an object moving through a medium, a concept that is used in many fields of physics. Most importantly, it requires relatively little extra work beyond what is normally done in the falling coffee filter lab. Thus a measurement of the added mass is a valuable addition to the standard coffee filter lab.

Table 1.  The added mass constant $C_A=m_A/(\rho R^3)$ measured for the falling coffee filters, and as calculated for various geometries moving through an ideal fluid.

| Conditions | $C_A$ |
|---|---|
| Measured in this experiment ||
| Coffee filter | $3.8 \pm 0.3$ |
| Theoretical values for an ideal fluid ||
| Coffee Filter | 3.40 |
| Disk [3,4] | 2.667 |
| Sphere [3,4] | 2.094 |
| Hollow sphere (outside plus inside) | 6.283 |

**Figures.**

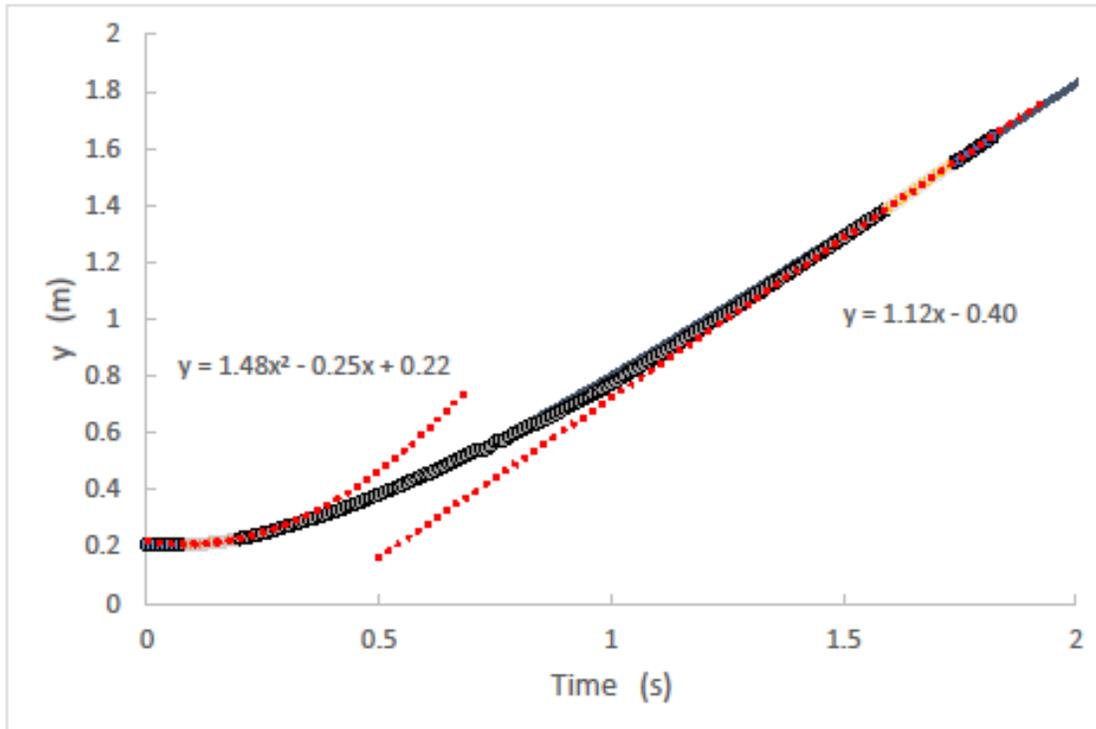

Fig. 1. Measured position versus time data for a single falling coffee filter (dark and light data points). The dotted lines (and equations) are the fits to the light data points. The solid curve (most of which is hidden under the data points) is the analytical solution in Eq. (6) using $A = 1.6$ m/s$^2$ and $V_T = 1.06$ m/s.

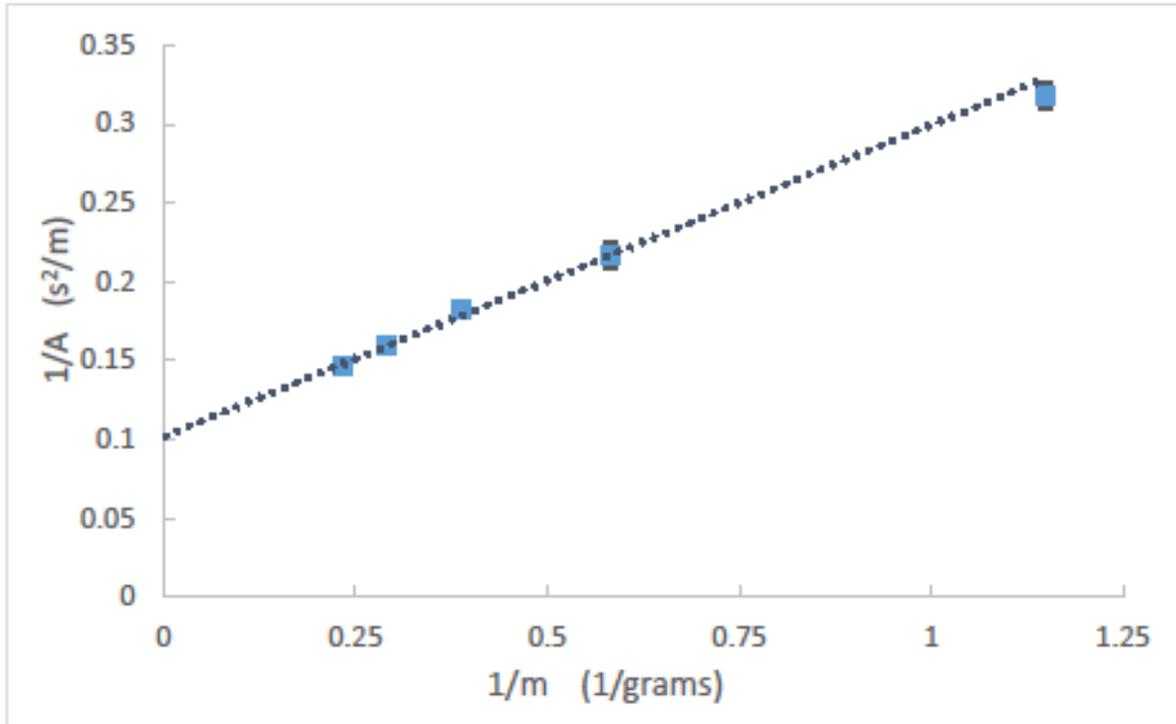

Fig. 2. Plot of 1/(initial acceleration) versus 1/(mass) of the falling coffee filters. The line is the prediction using Eq. (7) with $g = 9.82$ m/s$^2$ and $m_A = 1.9$ grams. There error bars are mostly smaller than the marker size.

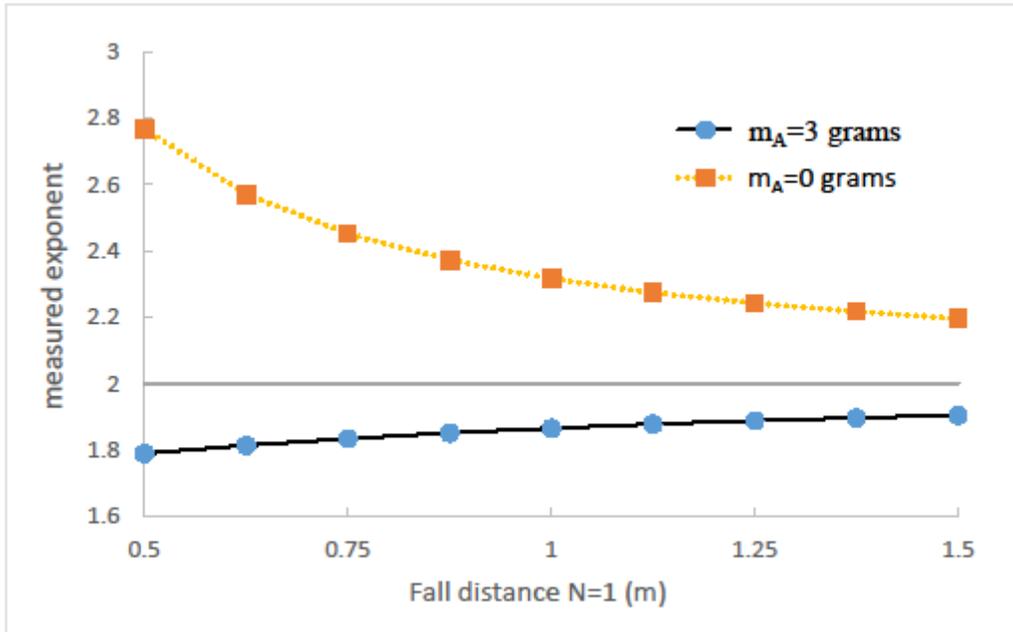

Fig. 3. The drag force's velocity exponent, $n$, that will be measured when observing falling coffee filters with only a measuring stick or tape measure, the "no tech" method. $m_A=3$ grams is the value that best fits the entire fall motion.

Supplementary Material for **The Added Mass of a Falling Coffee Filter** by C. Boyle and J. Pantaleone.

This supplement contains additional analysis of the data and details on the derivations.

**Extracting the scaling exponent, *n*, from the 40 Hz data.**

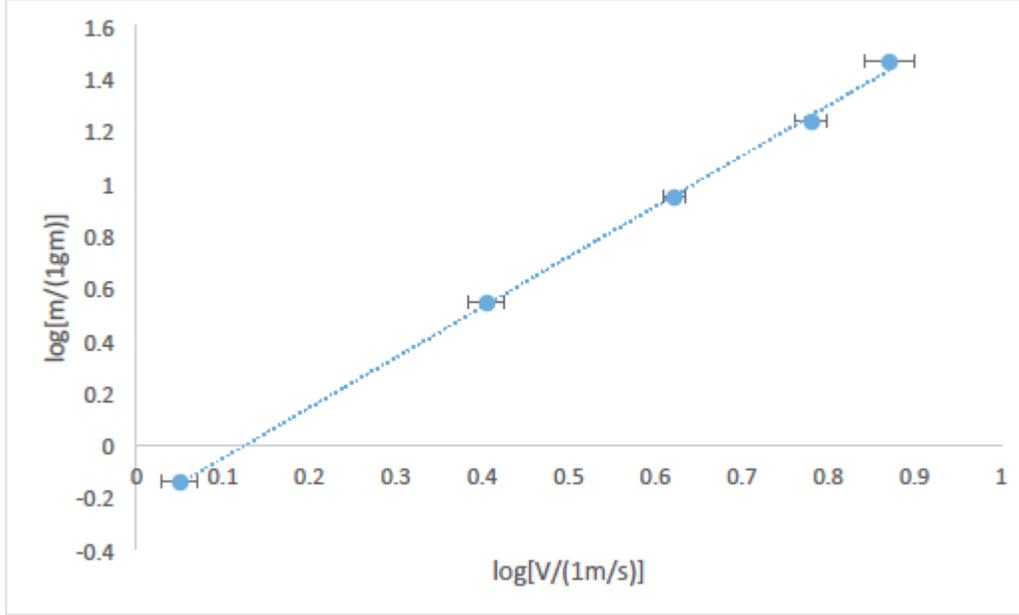

Fig. S1. Plot of log(mass) versus log(terminal velocity) where the terminal velocity was determined from measurements in the range 1.5 m < y < 1.8 m. The circular points are the average values and the statistical variation is given by the error bars. The dotted line is a linear fit to the data and has a slope of $n = 1.93 \pm 0.04$.

**Derivations for the equations in Section 4.**

Equation (11) gives the distance the coffee filter must fall from rest to be at 95% of the terminal velocity, $y_{95\%}$. This quantity can be found by first finding the time, *t*, for this to happen from the formula for velocity, *V*, as a function of time. This formula can be obtained by either integrating Eq. (3) or by taking the derivative of Eq. (6) to get

$$V(t) = V_T \operatorname{Tanh}\left[\frac{A}{V_T}t\right] \quad (S1)$$

where *A* and $V_T$ are as defined in the text. Setting $V(t)/V_T = 0.95$ and solving Eq. (S1) for the dimensionless time gives $At/V_T = 1.832$. Substituting this value into Eq. (6) gives the top line of Eq. (11)

$$y_{95\%} = \frac{V_T^2}{A}1.16 \quad (S2)$$

The parameters $V_T$ and $A$ depend on the mass of the coffee filters as described by Eqs. (4) and (5) in the text. The mass dependence can be turned into dependence on $N$, the number of coffee filters, by using $m=m_1N$ where $m$ is the mass of the coffee filters and $m_1$ is the mass of one coffee filter. Then Eqs. (4) and (5) can be rewritten as

$$V_T \approx V_{T1}\sqrt{N} \tag{S3}$$

and

$$A = \frac{g}{1+\dfrac{m_A}{m_1 N}} \tag{S4}$$

Here $V_{T1}$ is the terminal velocity of 1 coffee filter. Substituting Eqs. (S3) and (S4) into Eq. (S2), and evaluating the equations using the parameter values given in the text, yields the bottom line of Eq. (11) in the text.

Equations (12) and (13) follow from examining how the measured fall time, $t$, or fall distance, $y$, depend on $N$.

Analyses of the terminal velocity of coffee filters usually assume [1,2] that the steady-state drag force has the form $kV^n$, where $n$ is the scaling exponent and $k$ is a constant. Then at terminal velocity the drag force is balanced by the gravitational force

$$mg = kV_T^n \tag{S5}$$

The "low tech" and "no tech" experimental methods both neglect the approach to terminal velocity by taking the average velocity to be the terminal velocity.

$$V_T \approx V_{\text{averge}} = \frac{y}{t} \tag{S6}$$

They then analyze how the average velocity depends on $N$, the number of coffee filters, to find the scaling exponent, $n$. The error caused by using the average velocity can be calculated using Eq. (6). Working in the limit of large times, $At/V_T \gg 1$, then Eq. (6) has the approximate form

$$y(t) \approx V_{T1}\sqrt{N}\,t - \frac{V_{T1}^2}{g}\ln(2)\left[N + \frac{m_A}{m_1}\right] \tag{S7}$$

The first term in Eq. (S7) describes motion at terminal velocity according to Eq. (6) (and in agreement with Eq. (1)), while the second term gives the leading correction to that motion. The "low tech" and "no tech" methods differ in what they keep constant and what they vary, so the application of these equations to them must be discussed separately

The "low tech" method uses a fixed fall distance, $y$, and studies how the fall time, $t$, varies with $N$. Substituting Eq. (S6) into Eq. (S5), it is apparent that with $y$ a constant the scaling exponent can be found by constructing a $\log(t)$ versus $\log(N)$ plot and that ($-1/n$) will be the slope. Thus

$$\frac{1}{n} = -\left(\frac{N}{t}\right)\frac{dt}{dN} \tag{S8}$$

It is straightforward algebra to solve Eq. (S7) for $t$, and substitute this expression into Eq. (S8) to find $n$. Using that the correction term is small compared to the leading order term, the results simplify yielding Eq. (12) in the text.

For the "no tech" method, the fall time, $t$, is kept constant and the scaling exponent is extracted from plots of $\log(y)$ versus $\log(N)$. Then the scaling exponent is given by

$$\frac{1}{n} = \left(\frac{N}{y}\right)\frac{dy}{dN} \tag{S9}$$

Eq. (13) in the text is obtained by substituting Eq. (S7) into Eq. (S9), and simplifying the expression by using that the correction terms are small compared to the leading order term and that at leading order $t = y_1/V_{T1}$, where $y_1$ is the fall distance for $N = 1$.

In addition to the errors discussed here, both the "low tech" and the "no tech" methods suffer from the additional error that the speed of coffee filters decreases when they are near the floor. The effects of this error are not included in the above analysis.

**Calculating the air density.**

To compare the measured added mass to the value for an ideal fluid, the air density is needed. The air density depends on the atmospheric pressure, temperature and relative humidity. There are several online web pages that will calculate the air density from these inputs. An explanation of the calculation, along with all necessary parameters, can be found on Richard Shelquist's web page at https://wahiduddin.net/calc/density_altitude.htm